%% file: 1main_CameraReady.tex
\documentclass[lettersize,journal]{IEEEtran}
\usepackage{cite}
\usepackage{multirow}
\usepackage{graphicx}
\usepackage{amssymb,amsmath,times,wasysym}
\usepackage{pb-diagram}
\usepackage{float}
\usepackage{psfrag,balance}
\usepackage{srcltx,verbatim} 

\usepackage{fixltx2e}
\usepackage{multicol}
\usepackage{color}
\usepackage{algorithmic}
\usepackage{algorithm}
\usepackage{hyperref}
\usepackage{url}
\usepackage{caption}
\usepackage{subcaption}
\usepackage{tikz}
\usetikzlibrary{shapes.geometric, arrows, positioning} 
\tikzstyle{startstop} = [rectangle, rounded corners, minimum width=3cm, minimum height=0.5cm, text centered, draw=black, fill=blue!30]
\tikzstyle{io} = [trapezium, trapezium left angle=70, trapezium right angle=110, aspect=3, text centered, draw=black, fill=green!30]
\tikzstyle{process} = [rectangle, minimum width=1cm, minimum height=0.5cm, text centered, draw=black,fill=red!30]
\tikzstyle{decision} = [diamond, aspect=5, text width=2.5cm, text centered, draw=black,fill=orange!30]
\tikzstyle{arrow} = [thick,->,>=stealth]
\usepackage{tikz}
\usetikzlibrary{shapes.geometric, arrows, positioning} 
\tikzstyle{startstop} = [rectangle, rounded corners, minimum width=3cm, minimum height=1cm, text centered, draw=black, fill=blue!30]
\tikzstyle{io} = [trapezium, trapezium left angle=70, trapezium right angle=110, aspect=3, text centered, draw=black, fill=green!30]
\tikzstyle{process} = [rectangle, minimum width=3cm, minimum height=1cm, text centered, draw=black,fill=red!30]
\tikzstyle{decision} = [diamond, aspect=5, text width=2.5cm, text centered, draw=black,fill=orange!30]
\tikzstyle{arrow} = [thick,->,>=stealth]

\usetikzlibrary{shapes.geometric, arrows.meta, positioning, calc}

\tikzset{
	block/.style = {rectangle,minimum width=3cm, minimum height=.7cm,, draw, text width=9em, text centered, minimum height=3em},
	inputblock/.style = {block, fill=blue!30},
	sharedblock/.style = {block, fill=green!30},
	taskblock/.style = {block, fill=orange!30},
	lossblock/.style = {block, fill=red!30},
	avglossblock/.style = {block, fill=yellow!30},
	optblock/.style = {block, fill=cyan!30},
	updateblock/.style = {block, fill=purple!30},
	dashedarrow/.style = {thick,dashed, -Latex}
}

\newcounter{appendx}

\def\BibTeX{{\rm B\kern-.05em{\sc i\kern-.025em b}\kern-.08em
    T\kern-.1667em\lower.7ex\hbox{E}\kern-.125emX}}
\usepackage{balance}
\begin{document}
\bstctlcite{IEEEexample:BSTcontrol}
\title{ Frequency Range 3 for ISAC in 6G: \\Potentials and  Challenges}

\author{Gayan Aruma Baduge,  Mojtaba Vaezi, Janith K. Dassanayake,  Muhammad Z. Hameed, \\ Esa Ollila, and  Sergiy A Vorobyov
	
        
      \vspace{-5mm}  }

%

\maketitle

\begin{abstract}

Spanning 7–24 GHz, frequency range 3 (FR3), is a key enabler for next-generation wireless networks by bridging the coverage of sub-6 GHz and the capacity of millimeter-wave bands. Its unique propagation characteristics, such as extended near-field regions and spatially nonstationary fading, enable new transmission strategies. This article explores the potential of FR3 for integrated sensing and communication (ISAC), which unifies wireless communication and environmental sensing. We show that FR3’s bandwidth and multiple-input multiple-output (MIMO) capabilities enable high-resolution sensing, multi-target tracking, and fast data transmission. We emphasize the importance of ultra-massive MIMO with extremely large aperture arrays (ELAAs) and the need for unified near-field and far-field channel models to support efficient ISAC. Finally, we outline challenges and future research directions for ELAA-based ISAC in 6G FR3.

\end{abstract}

 \begin{IEEEkeywords}
 FR3, near-field, far-field,  integrated sensing and communication, ISAC, sub-6 GHz, millimeter wave, 6G.
\end{IEEEkeywords}

\section{Introduction}

Sixth-generation (6G) wireless technologies are expected to integrate communication and sensing capabilities within a unified  framework. 
Integrated sensing will not only enhance network performance but also enable new applications such as autonomous driving, indoor localization, and environmental monitoring and imaging. 
 A key design goal of \textit{integrated sensing and communication} (ISAC) \cite{Liu2022,Liu2018,vaezi2025ai} is the development of unified waveform and beamforming strategies that jointly serve both sensing and communication, 
 improving spectral, energy, and cost efficiency. These integration gains, however, critically depend on the optimal design of such waveforms and beamformers, tailored to the distinct propagation characteristics of the communication and radar channels within the frequency bands envisioned for 6G.

\begin{figure}[!t]
	\centering
	\def\svgwidth{250pt} 
	\fontsize{5}{2}\selectfont \vspace{0mm} 
	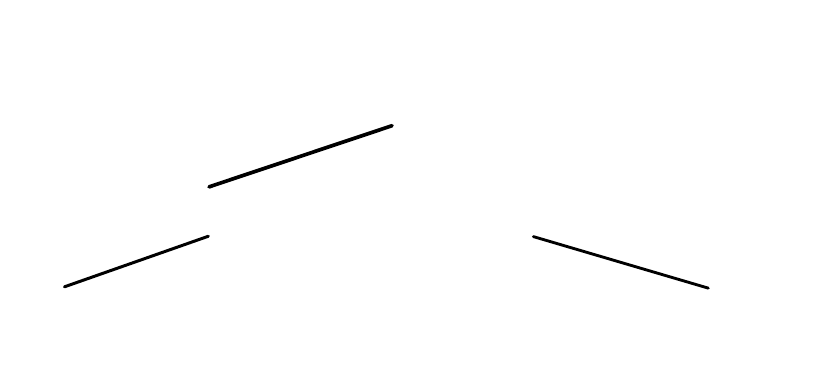 
    \caption{Spectrum allocations for FR3 alongside current FR1 and FR2 cellular bands.
    }
	\label{fig:FRs} \vspace{0mm}
\end{figure}

The \textit{upper mid-band frequency range}, spanning $7.125\,\mathrm{GHz}$ to $24.25\,\mathrm{GHz}$, also referred to as \textit{frequency range 3 (FR3)} or centimeter-waves (cmWaves), has been identified as a new band for 6G networks \cite{FCC2023,Bjornson2024,Kang2024,Bazzi2025,Cui2023}. This range bridges the gap between the previously defined FR1 and FR2 bands used in earlier generations of cellular networks, as illustrated in Fig.~\ref{fig:FRs} and defined below.
\begin{itemize}
\item \textbf{FR1:} $410\,\mathrm{MHz}$–$7.125\,\mathrm{GHz}$ (used in 2G to 5G; includes sub-6 GHz bands)
\item \textbf{FR2:} $24.25\,\mathrm{GHz}$–$71\,\mathrm{GHz}$ (used in 5G millimeter-wave)
\item \textbf{FR3:} $7.125\,\mathrm{GHz}$–$24.25\,\mathrm{GHz}$ (fills the gap between FR1 and FR2; proposed for 6G)
\end{itemize}

But why should a new frequency band be introduced for 6G? What are the implications, potentials, and challenges of such a choice in ISAC? We address these questions throughout this article. To put things into perspective, we first discuss the historical frequency band assignments used in previous generations of cellular networks.

For decades, the FR1 (which includes \textit{sub-6 GHz band}) has dominated mobile communications, including key frequencies used by 3G (850, 1900, and 2100 MHz), 4G (850, 1700/2100, 1900, and 2300 MHz), and 5G (3.3–3.8 GHz). The FR1 band offers several advantages, including wide-area coverage, strong signal penetration through buildings, and low hardware and deployment costs. However, FR1 suffers from severe spectrum congestion and lacks wide contiguous bandwidths, limiting its ability to support ultra-high data rates.

5G was designed to markedly enhance capacity, peak data rates, connectivity, and latency compared to 4G. In particular, it aimed to deliver up to $1000\times$ higher data volumes, $10$–$100\times$ more device connections, $10$–$100\times$ faster user data rates, and $5\times$ lower latency than 4G operating in FR1.
To achieve its ambitious data rates, 5G integrated advanced multi-user multiple-input multiple-output (MIMO) techniques with access to wide contiguous bandwidths  within the \textit{millimeter-wave (mmWave)} spectrum, more precisely in FR2 ($24.25\,\mathrm{GHz}$–$71\,\mathrm{GHz}$).

Despite attracting
significant research attention  over the past decade,  FR2 suffers from several major limitations, including reduced coverage, high penetration losses through buildings and foliage, and increased complexity and cost compared to FR1. A primary concern is that  mmWave signals struggle to penetrate obstacles such as walls, windows, and vegetation, unlike  FR1 signals. Moreover, the extremely narrow beams required by mmWave antenna arrays pose challenges for maintaining reliable connectivity under mobility. 
As such, mmWave has seen limited success as a mobile connectivity solution both in the U.S. and worldwide.  However, fixed wireless access based on mmWave has proven to be effective as an alternative to wired broadband in several use cases.  The global uptake of mmWave remains modest, as countries like South Korea and most of Europe have largely deprioritized mmWave due to deployment complexity and limited mobile use cases.

By offering wider contiguous bandwidth than FR1 and more favorable propagation characteristics than FR2, FR3 mitigates several key limitations of the existing cellular bands and has been identified as a strong candidate for 6G \cite{FCC2023,Bjornson2024,Kang2024,Bazzi2025,Cui2023,Miao2025}.
Referred to as the “golden band,” FR3 provides substantially larger continuous bandwidth for terrestrial mobile communications while enabling smaller antenna-array aperture sizes compared to FR1. 
It also benefits from lower path and penetration losses, reduced hardware complexity, and lower deployment costs relative to FR2 \cite{FCC2023,Bjornson2024,Kang2024,Bazzi2025,Cui2023,Miao2025}. 
Nokia’s recent coverage study~\cite{Nokia2024} shows that the lower FR3 band (7\,GHz) can achieve comparable or even superior cell-edge throughput to 5G at 3.5\,GHz (FR1) when using larger antenna arrays. 
The upper FR3 band (13\,GHz) supports reasonable coverage using directional antennas and additional transmit chains at the user equipment (UE) and customer premises equipment (CPE)~\cite{Nokia2024,Miao2025}.

As the bandwidth allocated to 6G in FR3 is limited—unlike the abundant bandwidth available in FR2—a natural solution is to exploit \textit{ultra-massive MIMO}   (UM-MIMO) by deploying a very large number of antennas. This leads to the concept of \textit{extremely large aperture arrays (ELAAs)}. Antenna aperture is the effective physical area over which an antenna or array receives or radiates electromagnetic energy. Larger apertures provide higher spacial resolution, gain, and sensing accuracy, and UM-MIMO provides aggressive spatial multiplexing gains for communication tasks \cite{Ramezani2024}.
ELAA-aided base stations (BSs) are thus expected to become a cornerstone for implementing ISAC in the FR3 band.

We argue that, due to the large physical aperture and high antenna count of ELAAs, the conventional \textit{far-field} channel assumptions used in MIMO systems (such as those in the 5G FR1) no longer hold. Hence, it is imperative to investigate \textit{near-field} channel models and techniques. The distinct propagation characteristics in the near-field will introduce new challenges and demand the development of novel transmission and  reception techniques.

This paper explores the potential of FR3 in enabling efficient ISAC for 6G. Our contribution can be summarized as 
\begin{itemize}
    \item \textbf{FR3 spectrum for ISAC:} We discuss available spectrum and challenges, including co-channel interference with existing satellite services, and highlight the need for agile spectrum access and management techniques.
    
    \item \textbf{UM-MIMO with ELAAs for ISAC:} We examine the need for  UM-MIMO with ELAAs to realize ISAC gains in FR3 and demonstrate how its unique propagation can be leveraged to provide seamless coverage along with high spectral and energy efficiency.
 
    \item   \textbf{ISAC implementation in FR3:} We emphasize the need for unified    far-field and near-field channel models that account for angle and distance dependencies.  
\end{itemize}

\section{The golden band (FR3) and its existing services}

Following the World Radiocommunication Conference 2023 organized by the International Telecommunication Union, specific FR3 sub-bands were identified as 6G candidate frequencies \cite{Bjornson2024}. In Region 1 (Europe, Africa, Mongolia, and the Middle East west of the Persian Gulf), the candidate band is 7.75–8.4 GHz; in Regions 2 and 3 (the Americas, most of Asia, and Oceania), it is 7.125–8.4 GHz. Additionally, the 14.8–15.35 GHz band was identified for use across all three regions. In Region 2, U.S. service providers could access up to 1825 MHz of spectrum for 6G deployment.

\subsection{Incumbent Communication and Sensing Services in FR3}

Spectrum regulatory agencies worldwide have identified numerous existing civilian and government services operating in FR3. These incumbent services include fixed wireless systems, space research, radio astronomy, aeronautical and maritime radio navigation,  non-terrestrial networks such as meteorological and Earth exploration satellites, maritime mobile and broadcast satellites, unmanned aerial systems, satellite feeder links, and broadband satellite Internet services.

According to the Federal Communications Commission, the largest spectrum allocations in the lower portion of FR3 (7.125–8.4 GHz) are held by fixed wireless and fixed/mobile satellite services for federal use \cite{FCC2023}. 
For example, the Federal Aviation Administration operates microwave links in this band to reliably connect air traffic control centers with remote  aeronautical radionavigation radars. Additionally, defense satellite communications systems use geostationary satellites operating in this range, with 7.25–7.7 GHz allocated for downlink and 7.9–8.4 GHz for uplink of defense and wideband gapfiller satellites.
Non-federal uses of the lower FR3 band include unlicensed allocations for ultra-wideband devices operating in the 7.75–8.75 GHz range. These devices serve several consumer sectors, including personalized item tracking, wall-penetrating radars, automotive radars, and wearable technology.

The upper portion of FR3 (8.5–13.75 GHz) hosts several incumbents, with the largest allocation for radiolocation services \cite{FCC2023}. 
Bands within 12.2-13.75 GHz may be available for spectrum sharing with full and limited access. Non-federal uses include the downlink and uplink of commercial fixed/mobile satellite services. Earth stations and user gateways use these bands for broadband Internet, serving enterprises, government entities, emergency responders, schools, and hospitals. For example, Starlink primarily operates in Ku (12–18 GHz) and Ka (26.5–40 GHz) bands, using 10.7–12.7 GHz for downlink and ~14 GHz for uplink. Federal uses include NOAA’s geostationary and data relay satellites, NASA’s active sensors, precipitation missions, and radars, and NSF-supported radio astronomy. Internationally, FR3 supports X-band (8–9 GHz) radars for civil and defense applications, land/naval/military radars, and fixed/mobile satellite services.

\subsection{Coexistence  of 6G ISAC Functionalities with Incumbents}

A key challenge in adopting FR3 for ISAC in 6G lies in addressing regulatory considerations, particularly enabling spectrum sharing and managing co-channel interference through advanced antenna arrays and signal processing.
As mentioned, many government and commercial communication, satellite, radar, and  navigation services operate in both the lower and upper portions of FR3 for mission-critical applications.

Terrestrial ISAC systems can coexist with these satellite services without significant performance degradation by employing advanced spatial interference cancellation techniques.
To enable the coexistence, ISAC  BSs  must employ unified waveforms and beamformers that jointly optimize communication, sensing, and spectrum sharing by maximizing sum rate and sensing accuracy under dynamic constraints, such as spatial interference nulling and permissible interference levels at satellite UEs and CPEs to ensure compatibility within FR3.

\subsection{Challenges of  Unification and Coexistence of ISAC in FR3}

The ISAC unification is constrained by the coexistence with a myriad of incumbents, particularly  including  satellite services in FR3  and underlying optimization procedures are also challenging. This is because the  design objectives of  communications, sensing, and spectrum sharing tasks are mutually coupled and competing each other.
To meet data rates envisioned for 6G, innovation of UM-MIMO technologies in FR3  will also be needed  \cite{Bjornson2024}.    
Classical model-based multi-objective optimization  and   signal processing techniques may sometimes become prohibitively complicated for   designing 6G ISAC with incumbents in FR3 due to fundamental design trade-offs \cite{vaezi2025ai}.  The optimal solutions may  not always be feasible due to model inadequacy, or optimal solutions may also be too complicated for real-time implementation.
 In such cases, data-driven learning-based  end-to-end design approaches may   be appealing and can supplement the model-based designs \cite{vaezi2025ai}.

\begin{figure}[!t]\centering
	\def\svgwidth{240pt} 
	\fontsize{8}{4}\selectfont 
	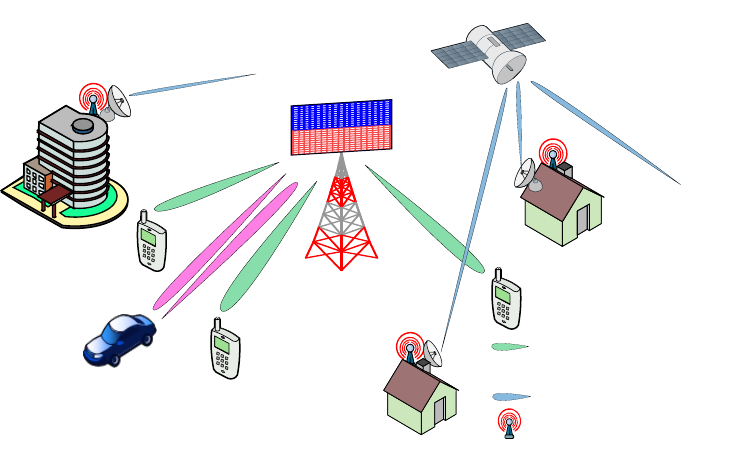
	\caption{Coexistence of UEs and satellite CPEs in FR3 with a BS using an ELAA.
 }\label{fig:CPE_UE_Coex}
\end{figure}

The above challenges are compounded by the need for accurate sensing of satellite CPEs operating in the FR3 band to enable ISAC coexistence (see Fig.~\ref{fig:CPE_UE_Coex}). Co-channel interference must be efficiently managed, which requires spatial interference nulling toward satellite CPEs while preserving ISAC functionality. This, in turn, demands precise CPE localization, dynamic spectrum sensing, and channel estimation. While classical pilot-based techniques can estimate BS-UE channels, passively sensing satellite CPEs is extremely difficult due to their highly directional links.

Passive sensing of satellite CPEs appears practically unrealizable, even for modern BSs, but active sensing is feasible through pilot-based UE-BS channel estimation. A pragmatic solution involves beacon/tone-based active sensing, where satellite CPEs periodically transmit omni-directional beacons. With mutual agreements between 3GPP and satellite providers, ISAC-enabled BSs can use these beacons for accurate localization, angle-of-arrival estimation, and dynamic spectrum sensing. Efficient sensing of satellite CPEs in FR3 for ISAC coexistence remains an open research challenge.

\section{Candidate technologies for   6G ISAC in FR3}

This section elaborates candidate technologies to overcome the key challenges of designing and  implementing ISAC with incumbent services in FR3.

 \subsection{ELAAs for Enabling ISAC in  FR3}

Co-located massive MIMO technology is currently  serving as  one of the key enablers of  5G  deployments. The 5G BSs operating in lower FR1 band  (3.5\,GHz)  are typically equipped with   16, 32  or 64   antenna arrays.
They adopt \textit{time-division duplexing}  with full digital processing such that each antenna is connected to a separate RF chain.  Since mmWave 5G is not suitable  for extensive coverage, it is typically used as a supplementary service in dense urban areas with high traffic. Aggressive spatial multiplexing is not typically  needed for  mmWave  thanks to large bandwidth availability in this band, and  high data rates can still be attained through efficient beam-searching and   beam-steering techniques. Fixed wireless access is also a major use-case of   5G mmWaves. The apertures   of  arrays used for 5G mmWaves are  much smaller than those used for  the  5G sub-6\,GHz band for the same number of antennas.

There are two proven techniques to linearly increase the  data rates, i.e.,  \begin{enumerate}
    \item using   wider bandwidths in unoccupied spectrum,  and 
    \item leveraging spatial multiplexing gains rendered by large antenna arrays.
\end{enumerate} The first  technique is currently being  utilized by 5G mmWave BSs to boost data rates as there is an ample amount of usable  bandwidths within FR2. This is because the achievable rate ($R$) is proportional to the communication bandwidth $(W)$ due to Shannon's channel capacity theorem: $R = W\log_2(1+\gamma)$, where $\gamma$ is the signal-to-noise ratio (SNR).  However, the propagation in FR2 is not favorable to provide seamless coverage and connectivity for cellular services, and this band is being used mostly for fixed wireless access.  Most 5G cellular systems around the world have been deployed in FR1, and total bandwidth allocation for 5G in this sub-6\,GHz band is about 1200\,MHz  (see Fig. \ref{fig:FRs}). 

Interestingly, the total bandwidth availability for 6G in FR3 is comparable to that of 5G in FR1. While the maximum allocation in FR3 reaches 1825 MHz across all regions, Region~1 (including the U.S.) is allocated only 1200 MHz. Thus, relying solely on wider bandwidths may be insufficient to meet 6G data rate targets, given the similar overall bandwidth availability.

 The second technique   to linearly increase the  data rates involves transmitting  multiple data streams to serve many users in the same time-frequency resource element by using spatial multiplexing with antenna arrays.      This  technique is particularly appealing to increase the   spectral efficiency ($\eta$),  defined as the data rate per unit bandwidth. Specifically, $\eta$ is a  linearly increasing function of the spatial multiplexing rate, $\min(M,K)$, where $M$ is the number of  antennas at the BS and  $K$ is the sum of antennas at all UEs. Increasing the number of antennas, especially at the BS, is thus a practical way to boost data rates for a given bandwidth.

Packing an enormous number of electrically-small antennas  gives rise to ELAAs  \cite{Ramezani2024}. Unique propagation characteristics of large antenna arrays such as  ultra-high spatial resolution, super-directivity, and finite spatial depth-of-foci can be leveraged to improve the performance of sensing tasks, including localization, target detection, and parameter estimation. We henceforth categorize ELAA-aided wireless communication to the broad category of    UM-MIMO, which will be the foundation of implementing ISAC in 6G within FR3.

\vspace{-3mm}

\subsection{Challenges of Implementing ISAC with ELAAs in FR3}

\begin{figure*}[!ht]
  \centering
  \begin{subfigure}[b]{0.45\textwidth}
\includegraphics[width=0.98\textwidth]{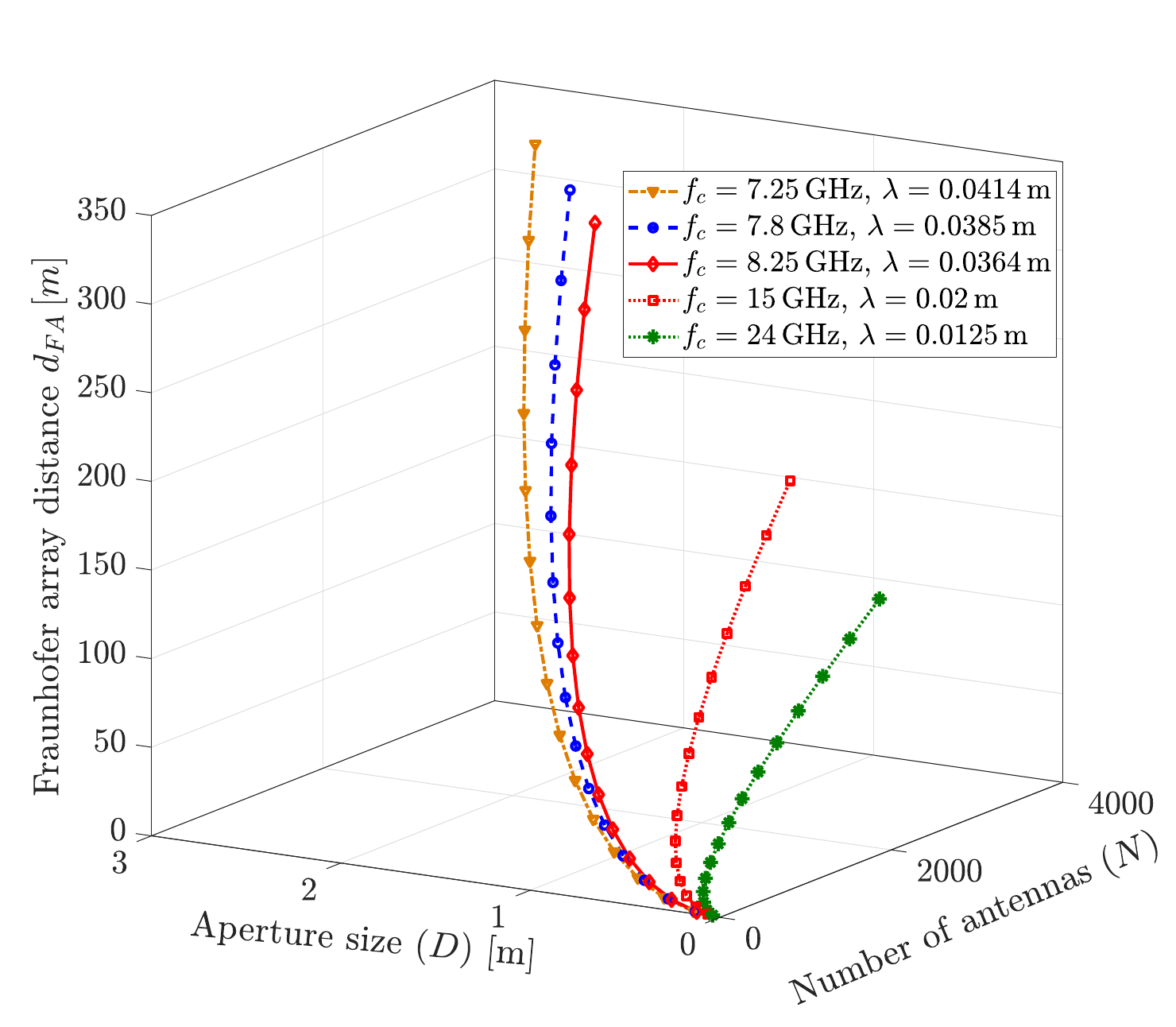}
	\caption{Fraunhofer distance versus ELAA  aperture size  and number of antennas in  FR3 band.   
    }
	\label{fig:dFArray_array_size_N}
  \end{subfigure}
  \hfill
  \begin{subfigure}[b]{0.45\textwidth}
   \includegraphics[width=0.98\textwidth]{	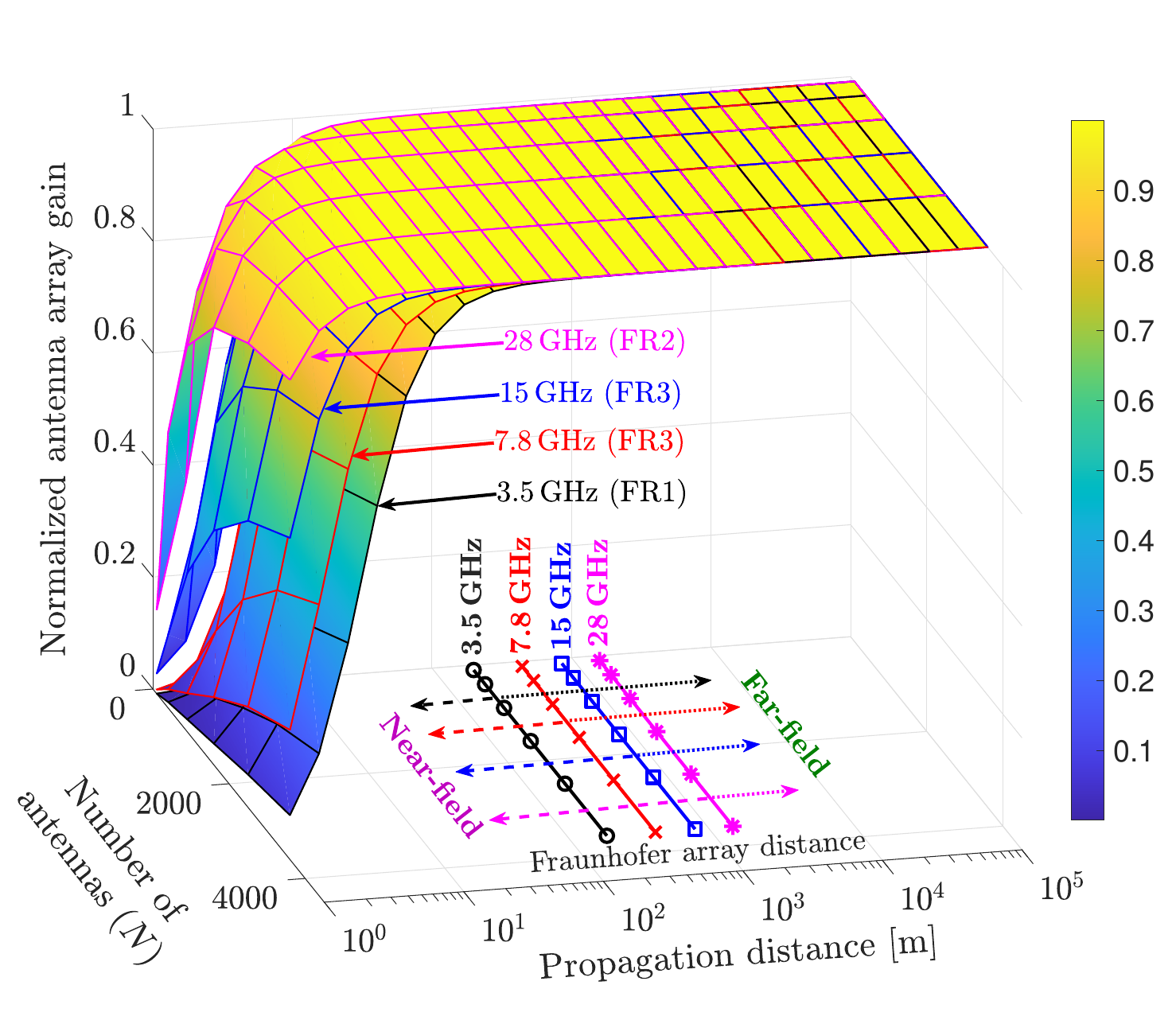}
	\caption{The normalized array gain against the propagation distance and number of antennas.}
    \label{fig:array_gain_prop_dis_N}
  \end{subfigure}
  \caption{Performance of ELAAs across different frequency bands.} \vspace{-4mm}
  \label{fig:combined}
\end{figure*}

Current MIMO systems for communication and radar services are based on two exclusive assumptions:
\begin{enumerate}
    \item [i.] all UEs and radar targets are located in the \textit{far-field} of the antenna array, and 
    \item [ii.]  the channels exhibit wide-sense stationary (WSS) properties in the spatial domain.
\end{enumerate}
Thus, existing designs assume that channel models with locally planar wavefronts over the antenna array are valid, and that the entire channel is visible to all UEs/targets.  The accuracy of these classical far-field channel models has been thoroughly verified for the FR1 band through both channel measurement campaigns and theoretical limits~\cite{Gao2015}. 

For all practical transmission distances in the FR1 band, it is guaranteed that almost all UEs/targets will be located in the far field of the antenna array. This is because the \textit{Fraunhofer array distance} ($d_{\mathrm{FA}}$), which demarcates the boundary between the radiative \textit{near-field} and \textit{far-field} regions of an antenna array, is on the order of a few meters. This follows from the expression $d_{\mathrm{FA}} = 2Nd_a^2/\lambda$~\cite{Ramezani2024}, where $d_a$ is the largest dimension (diagonal) of an antenna element, $N$ is the number of antennas in a planar array, and $\lambda$ is the operating wavelength.

The adoption of far-field channel assumptions for MIMO system designs is, therefore, well justified for 5G in the FR1 band, as the underlying antenna arrays are \textit{electrically} small (i.e., their dimensions are small relative to $\lambda$). However, in general, the same cannot be stated for 6G systems operating in the FR3 band due to fundamental changes in antenna array design and propagation characteristics.
The existing far-field models are valid for the FR3 band only when the UEs or targets lie beyond the Fraunhofer distance.

In Fig.~\ref{fig:dFArray_array_size_N}, the variation of $d_{\mathrm{FA}}$ with the aperture size of the ELAA ($D$) and the number of antenna elements $N$ is shown for five carrier frequencies in the FR3 band. The ELAA is modeled as a  square uniform  planar array (UPA)  consisting of $N$ tiny antenna elements, each with horizontal and vertical dimensions of $\lambda/2$. The spacing between adjacent elements is also $\lambda/2$, making $D$   proportional to $\sqrt{N}$. Figure~\ref{fig:dFArray_array_size_N} reveals that $d_{\mathrm{FA}}$, and hence the near-field region, can extend to hundreds of meters when ELAAs are deployed in the FR3 band.

Under the far-field or planar-wavefront assumption commonly used in prior work, the array response—including steering vectors in communications and target response matrices in radar—depends only on the angles of arrival and departure (AoA/AoD).
The far-field \textit{radar cross section} (RCS)   depends solely on  the extended target size and  wavelength
 \cite{Knott1993}.  The key implications of the spatially WSS  assumption are that the  statistical properties of the received signals/echoes, such as the average received power and AoDs/AoAs,  do not vary  over the array's spatial dimensions.  Via this WSS assumption, the following simplifications have been invoked in  the existing ISAC models and designs; (i)  the statistical mean of the  spatial channel coefficient    is a constant across the   array,   and (ii) the covariance  depends only on the difference between the antenna indices.  Another implication   is that   existing ISAC designs rely on  the fact that UEs and targets are visible to the entire array,   thus paving the notion of wholly-visible   scatterers  and full visibility regions  in   existing ISAC  channel   models \cite{Gao2015}.

 6G BSs in FR3 must use large co-located or distributed  ELAAs to achieve the high spectral and energy efficiency needed, as FR3 offers similar bandwidth to FR1 but suffers from higher path loss.
Previously, we observed that the Fraunhofer array distance is precisely $N$ times larger than the Fraunhofer distance of an individual antenna. Hence, the combination of ELAAs and  higher frequency bands    undermines the fundamental assumptions that  the wavefronts of radiated electromagnetic  waves are locally planar over  the antenna arrays and that the underlying channels are spatially WSS. 
Hence, the adoption of ELAAs in FR3  largely extends the radiative near-field of the  antenna array  due to increased aperture sizes and operating frequencies (see Fig. \ref{fig:dFArray_array_size_N}).
Consequently,      the UEs and sensing targets may be  located predominantly within the radiative near-field. Such wireless systems  exhibit fundamental changes in  electromagnetic properties and propagation characteristics. Wireless transmission and reception with  ELAAs must also be explored by using integrals over the   continuous apertures instead of the conventional approach of summations of individual antennas. 

One of the primary  purposes of ELAAs  is to adopt beamforming to
achieve a much larger total channel gain than that  with a single antenna element. With the far-field assumption, the effective  array gain typically increases linearly with the number of antennas, and it is independent of the propagation distance ($d_p$). Hence, the antenna  and array gains can be decoupled in the far-field.  However, the  gains of an antenna and the array  in the near-field must be treated jointly, and it exclusively depends on $d_p$.  Figure~\ref{fig:array_gain_prop_dis_N} shows the variations of the array gain of an ELAA  against  $d_p$ and $N$ in the array. Specifically,  the normalized array gain,  which is computed as a ratio between the total received power of the ELAA with  $N$ antennas and  the received power  of $N$ reference antennas, is plotted while keeping the arrays aperture size ($D$) the same for the  frequencies  3.5\,GHz in FR1,   7.8 and 15\,GHz in FR3, and  28\,GHz in FR2. Since $D$ is fixed, the Fraunhofer array distance increases with the carrier frequency.   It is seen that the normalized array gain varies with $d_p$ and $N$. This    justifies the need to capture unique near-field  propagation characteristics for the efficient implementation of ISAC in the FR3 band.

For ELAAs in the FR3 band, the UEs and radar targets for automotive applications  are   likely to be located in the extended  near-field.
Rigorous measurement campaigns  \cite{Gao2015} have revealed that   the  statistical signal properties  vary  over ELAAs, and the spatial WSS assumption may no longer be valid.  
Consequently,  the far-field planar wavefront  and spatially WSS assumptions may no longer be valid for ELAAs. Thus,  the existing physical-layer  ISAC designs  based on  far-field channel models  will  become inaccurate/incomplete once ELAAs are adopted in  the FR3 band. 
The first key challenge is to establish theoretical foundations for unique ELAA channel features such as multiple depth-of-foci,  non-WSS and visibility regions in the spatial domain to implement  ISAC in the FR3 band. The second challenge is  to discover  efficient ISAC system designs that leverage  unique ELAA near-field  channel features to     practically realize integration gains of ISAC designs.

Super-directivity and favorable propagation characteristics of ELAAs in FR3  will also lead to finer spatial resolutions and significantly lower transmit power requirements \cite{Ramezani2024}. The combination of unprecedented spatial multiplexing gains and low   power requirements will pave the way to realize (i)  large spectral/energy efficiency gains for communication tasks and (ii) boost  estimation, detection, and localization  accuracy for sensing tasks in ISAC systems, subject to fundamental design trade-offs for 6G operating in FR3.

The extremely large apertures and propagation characteristics of the FR3 band demand novel precoding/combining and signal processing techniques. 
Key challenges include interference rejection, frequency-selective beamforming to address beam squint, and dynamic multi-user precoding to handle spatial non-stationarity and varying visibility regions across the array.
All the  above distinctions with respect to  massive MIMO in the FR1 band justify that  radically novel channel/signal models, signal processing techniques, and  new performance  analysis frameworks are necessary to explore the true potential of     ELAAs in FR3  as an enabling  ISAC technology  in 6G.

\begin{figure}[t!]
	\centering
	\includegraphics[width=0.45\textwidth]{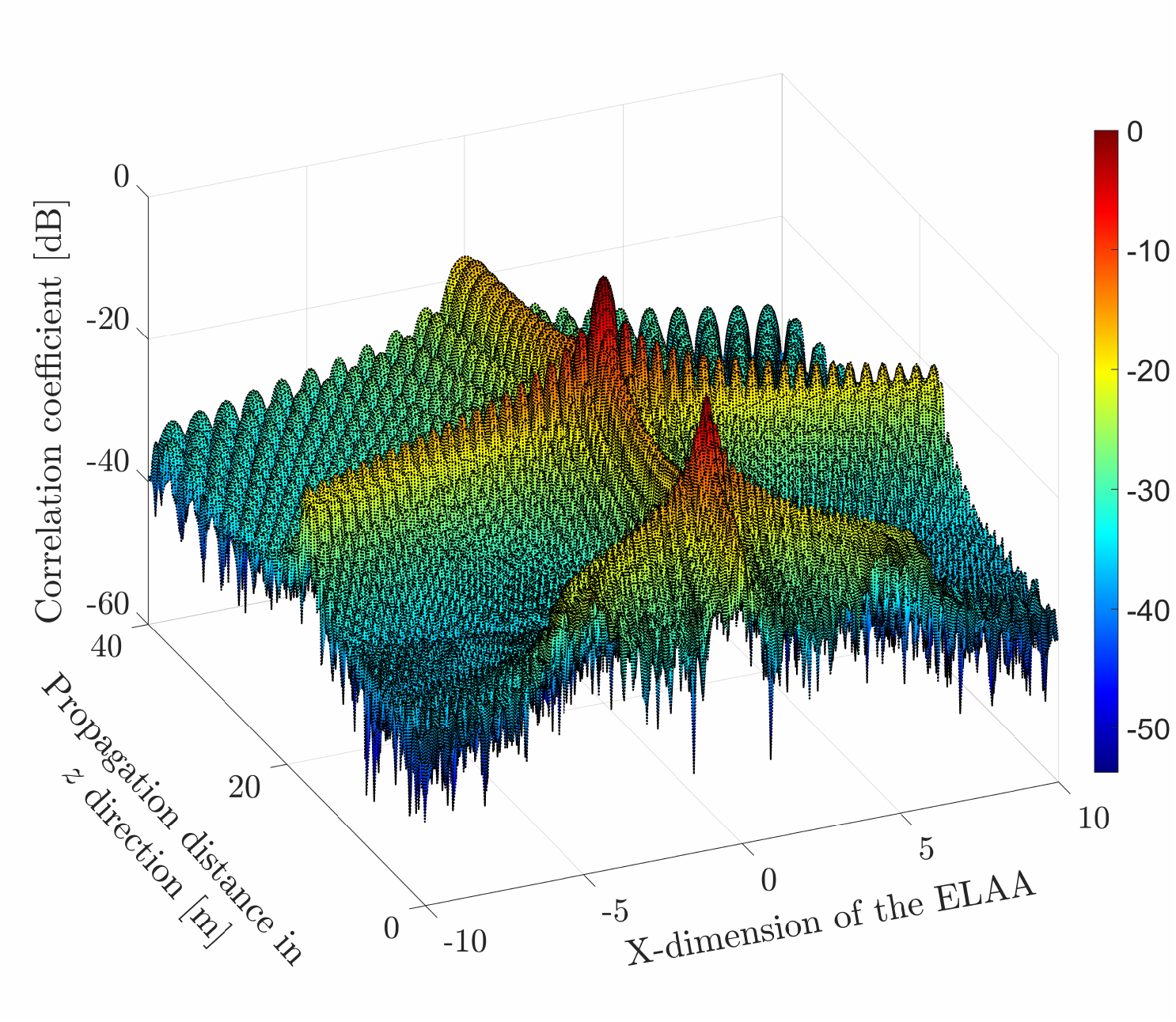}
	\vspace{-1mm}
	 		\caption{Correlation coefficient in dB scale
		when one UE is fixed and a second UE
	is moved along the same spatial direction.}
		\vspace{-5mm}
	\label{fig:Corr_coe_dB_15}
\end{figure}

\vspace{-3mm}

\subsection{Implications of  Extended Near-Field for  ISAC in FR3}

To efficiently implement ISAC in the FR3 band,  unified and generalized near/far-field channel models need to be developed. Unique ELAA propagation features must also be leveraged  to practically realize the integration gains of ISAC. To this end, the  ELAA can be modeled  as a  2D  transmit antenna surface with a densely-packed large number of  antennas, each having a finite area. These antennas are assumed to be placed edge-to-edge to  constitute  an extremely large contiguous antenna surface with an approximately continuous electrically-large  aperture.
Thus, the  propagation effects must be investigated  using 2D integrals over the entire aperture instead of typical summations of individual antennas. 
The channels between the ELAA  and UEs/targets must also be modeled via a superposition of spherical wavefronts by virtue of circular and divergent phase gradients to capture distinct propagation distances, effective antenna areas, and polarization mismatches in     spatial/angular domains.  The existing techniques for modeling and analyzing near-field ISAC have been reported  in \cite{Zhao2014}. For example, the near-field sensing channels employ a steering vector-based formulation, which depends not only on spatial angular signatures (AoDs/AoAs) but also on distance  due to   spherical wavefronts.

 New near-field ISAC channel models must  capture spatial finite depth-of-foci, non-WSS, small-scale, and large-scale fading effects. The array response vectors of both communication and radar channels   depend not only on AoAs/AoDs in  the elevation and azimuth planes   but also on the transmission   distances. 
 Figure~\ref{fig:Corr_coe_dB_15} shows  that the   array steering vectors  of two UEs located in the same  azimuth and elevation directions    correlate and decorrelate repeatedly along the distance domain in the near-field of ELAAs. Hence, the  beams transmitted by an ELAA possess multiple focal points with finite depths along the same spatial direction in the near-field as opposed to the infinite depth-of-focus of   beams in far-field channel models with plane-wavefronts.  
Multiple UEs/targets having the same spatial direction can hence be served/illuminated by using near-field beam-focusing. This task requires the design of unified beamformers and waveforms that can be jointly used for both communication and sensing functionalities     within the same bandwidth allocated in the FR3 band.

\begin{figure}[t!]
	 	\centering
	\includegraphics[width=0.44\textwidth]{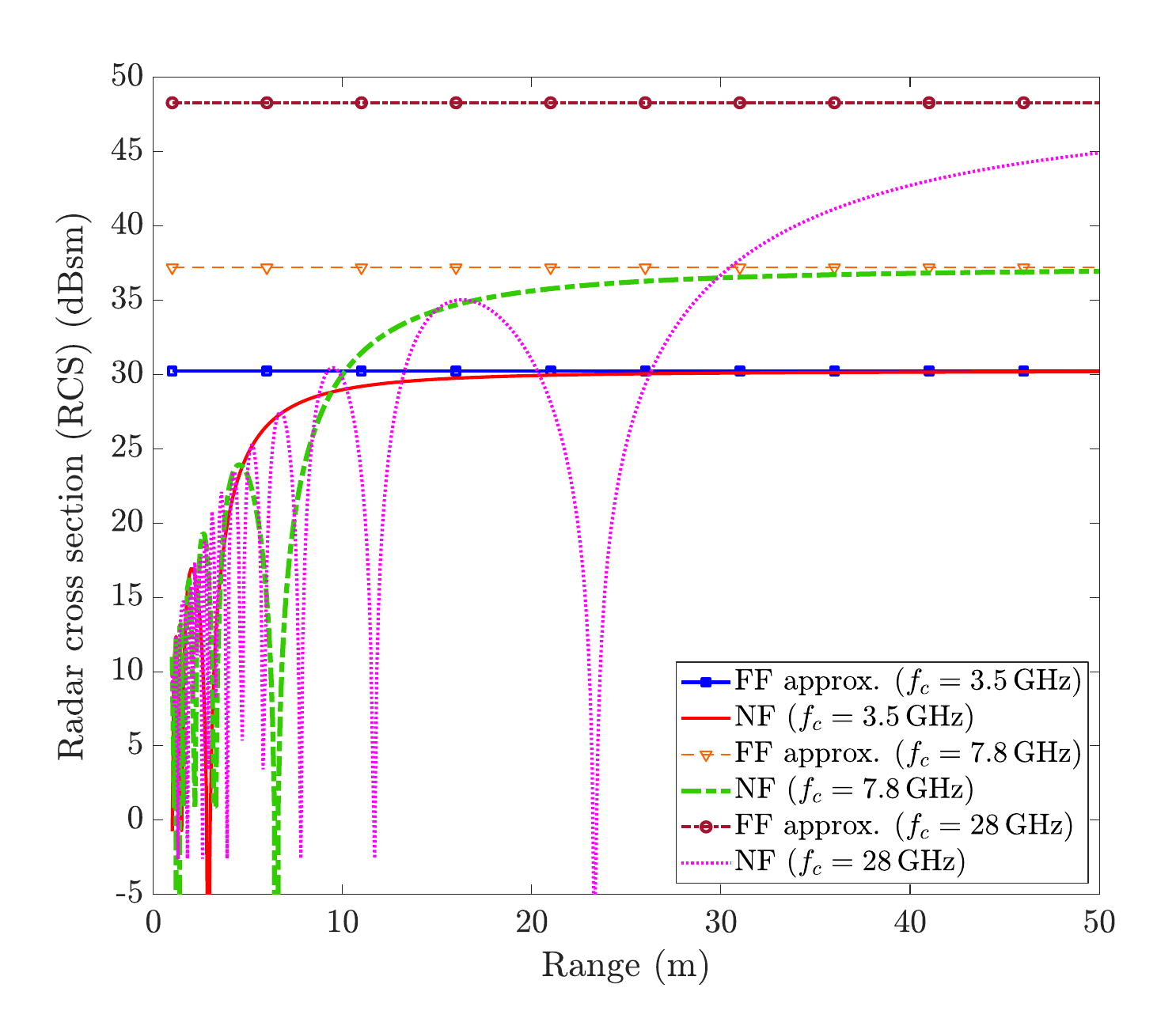}\vspace{-3mm}
	\caption{The RCS versus   range for different frequency bands with far-field (FF) and near-field (NF) assumptions.}\vspace{-5mm}
	\label{fig:RCS}
\end{figure}

The target response matrix of   a radar channel is a function of RCS and  the array steering vectors \cite{Knott1993}. 
The classical definition of the RCS of an extended target   exclusively relies on the far-field planar wavefront assumptions. Hence, the RCS becomes independent of the target range ($r$). For example, for a circular 2D target with a radius $a$, the RCS with the far-field definition is given by $\sigma    = 4\pi^3 a^4/\lambda^2$ \cite{Knott1993}. However, Fig. \ref{fig:RCS} uncovers that  once the electric field in the near-field ELAA channels is considered, the  RCS will also depend on the range.  Hence, the   radar target response matrices in the near-field will  depend on the   range  in addition to the       AoAs/AoDs in  the  elevation and azimuth planes.
In the  far-field,  the  range and AoAs  are estimated based on the time-delay and target response matrix \cite{Liu2022}, respectively. In   state-of-the-art sensing systems, the resolution of  time-delay and time difference of arrival  estimations is typically hindered by the  bandwidth and synchronization requirements  \cite{Liu2022}.  
However, with accurate near-field channels for ELAAs in the FR3 band, both the range and AoA of a target  can  be estimated  directly through the range-angle dependent  target response matrix without relying on   traditional time-delay-based estimators.

Accurate sensing of unique ELAA channel characteristics in the extended near-field of FR3 is essential for enabling  ISAC integration gains through unified waveform and beamformer design, i.e., communication waveforms operating also in the delay-Doppler domain or combined communication/sensing waveforms. These characteristics can be captured via spatial-time, spatial-frequency, angle-delay, and delay-Doppler representations using 2D Fourier transforms. Delay-Doppler domain modeling is especially useful, as its parameters map directly to scatterers and allow embedding communication symbols in unified waveforms. ELAA's finite focal depths and high angular resolution enable novel multiplexing, allowing multiple UEs with similar angles but different ranges to be served simultaneously without interference \cite{Ramezani2024}. Additionally, ELAA super-directivity helps mitigate spatial aliasing, enhance radar resolution, and reduce clutter in 6G ISAC systems.

Near-field beamforming aims to tackle the interplay between near-field propagation and spatial-wideband effects, such as beam squint and split, in wideband ELAAs \cite{Ramezani2024}.
 True-time delays (TTDs) can be incorporated into hybrid beamformer architectures to mitigate the joint impact of spatial-wideband fading effects in the near-field.  
The candidate beamformer design approaches are:
\begin{itemize}
\item {\it Sub-array partitioning}: A large array is divided into several smaller sub-arrays, each approximated as operating in the far-field planar region. 
\item {\it Application integration}: Near-field beamforming is combined with other emerging technologies such as reconfigurable intelligent surfaces (RIS) and holographic meta-surface-based antenna arrays. For instance, RIS Fresnel zone properties can be exploited, with optimization adapted to TTD and analog beamforming constraints.
\item {\it TTD configuration}: Exploring different TTD configurations, including parallel, serial, and hybrid architectures.
\item {\it Deep learning-based methods}: This includes transformer-based and end-to-end learning-based methods,  and can address the real-time design challenges of UM-MIMO ISAC in FR3 by  supplementing the model-based classical iterative optimization algorithms with  learning-based solutions and  provide reduced computational complexity and robustness to model mismatches \cite{vaezi2025ai}.
\end{itemize}

 \vspace{-3mm}

 \section{A use-case of  FR3 for deploying ISAC in 6G}

  We investigate a use-case of implementing ISAC in the FR3 band with an ELAA-equipped BS.  For a fair comparison across the frequency bands, the antennas at  ELAA are  arranged as a square UPA of a fixed size $1.243 \times 1.243$\,m to keep  a fixed aperture size.   The size of an antenna element is $\lambda/4 \times \lambda/4$. The relative antenna spacing is set to $\lambda/2$. 
Two carrier frequencies in the FR3 band (7.8 and 15\,GHz) are used, and the FR1  frequency  3.5\,GHz  is used as a baseline comparison.
For the same aperture size, the numbers of antennas in the ELAA for 3.5, 7.8 and 15\,GHz 	are 400, 961, and 1521, respectively. 
  The UEs, target, and clutter lie in the near-field, with distances below the Fraunhofer limit ($d_{\text{FA}} = 72.1$\,m for $f_c = 3.5$\,GHz). For 7.8 and 15 GHz,  $d_{\text{FA}}$ limits are 160.65 m and 308.96 m, respectively. 
 The BS-UE  distances are set to $[30, 40, 50 , 60]$\,m. The extended target has four scatterers, and their distances are set to $[30,30,31,31]$\,m.  
 The coherence interval $\tau_c $ is set to 196, and the pilot  length for uplink channel estimation is set to $\tau_P=K$. The uplink pilot transmit SNR is set to  0\,dB.

\begin{figure}
 	\centering
 	\includegraphics[width=0.44\textwidth]{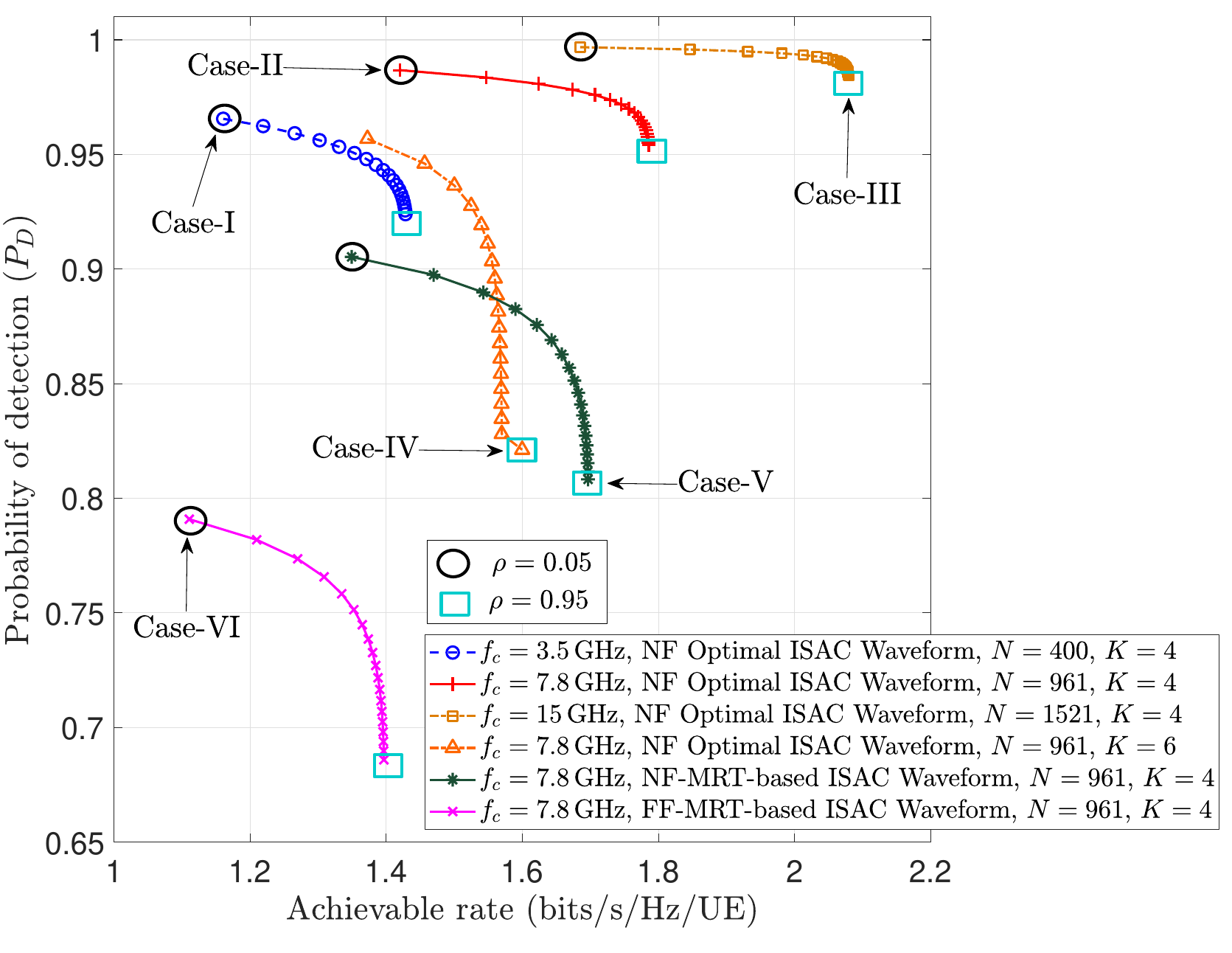}\vspace{-3mm}
 	\caption{Trade-off between the  rate and detection probability.}\vspace{-7mm}
 	\label{fig:Tradeoff}
 \end{figure}

In Fig. \ref{fig:Tradeoff}, the trade-off between the achievable rate per user ($\mathcal R$) and the probability of detection ($P_D$) is depicted for the  optimal ISAC waveform  designed for the near-field channel models for three carrier frequencies $f_c\in\{3.5, 7.8, 15\}$\,GHz.   The ELAA's aperture size  is kept fixed across all bands for a fair comparison.    The target detection is performed by adopting the detector proposed in \cite[Eq. (69)]{Khawar2015}, with a fixed false alarm probability of $10^{-7}$. Six rate-detection trade-off curves are generated by varying the numbers of UEs  and ELAA antennas as $K \in \{4, 6\}$ and   $N\in\{400, 961, 1521\}$, respectively. These curves are obtained by adjusting a weighting factor, $\rho$, which governs the priority assigned to communication versus sensing objectives in ISAC waveform design, as proposed in \cite{Liu2018}. Here, $\rho$ is varied from 0.05 to 0.95 in increments of 0.05 at the average  transmit SNR of 10\,dB.
Each point along the trade-off curves corresponds to a distinct $\rho$ value, indicating different ISAC operating points. Figure~\ref{fig:Tradeoff} illustrates the fundamental ISAC trade-off: as the communication rate $\mathcal{R}$ increases (higher $\rho$), the detection probability $P_D$ decreases, indicating a shift toward communication at the expense of sensing.

In Cases~I to IV, an optimal ISAC waveform is used, by jointly minimizing multi-user interference and the mismatch with a reference sensing waveform having good correlation properties~\cite{Liu2018}.
Since the ELAA's aperture size is kept fixed for all three frequencies, the number of $\lambda/4\times \lambda/4$ antennas that can be packed in the UPA increases with higher frequencies.   
Hence, it is evident that increasing the number of antennas in the ELAA enhances both the UE rates and detection probability.
An MRT-based ISAC precoder using an accurate near-field model (Case~V) is compared against a far-field-based design (Case~VI) to assess performance loss.
Comparing Cases~V and VI shows noticeable sum rate and detection losses when a far-field planar wavefront-based ISAC precoder is used when UEs are actually located in the near-field   at 7.8\,GHz (FR3). Comparing Cases~II and IV reveals degraded detection performance as the number of UEs increases.

\section{Conclusion}

We have investigated the potential of the FR3 for ISAC in  6G. Our work suggests that this ``golden band” offers a unique opportunity to achieve wide-area coverage, high data rates, and high-resolution sensing. Toward this end, we highlighted the importance of unified near- and far-field channel models, dynamic spectrum sensing,  and advanced transceiver designs for enabling seamless coverage and enhanced performance for ISAC in FR3. To fully exploit the propagation characteristics of the FR3 band for ISAC, the use of ultra-massive MIMO with ELAAs is a practically viable path forward.

\vspace{-3mm}
 \section*{Acknowledgment}
 This work has been supported in part by the NSF under Grants CCF-2326621 and CCF-2326622.

\vspace{-4mm}


\vspace{-14mm}
\begin{IEEEbiographynophoto}{Gayan  Aruma Baduge} 
is  a Professor in 
the School of Electrical,  Computer, and Biomedical   Engineering,  Southern Illinois University, IL, USA.	
\end{IEEEbiographynophoto}

\vspace{-14mm}

\begin{IEEEbiographynophoto}
{\bf Mojtaba Vaezi} is an Associate Professor in the ECE Department at Villanova University. He is the recipient of the 2020 IEEE Communications Society Fred W. Ellersick Prize.
\end{IEEEbiographynophoto}

\vspace{-14mm}

\begin{IEEEbiographynophoto}{Janith Dassanayake} is 
a Ph.D. student in the School of Electrical, Computer, and Biomedical  Engineering, Southern Illinois University,  IL, USA. 
\end{IEEEbiographynophoto} 
 
\vspace{-14mm}

\begin{IEEEbiographynophoto}{Muhammad Zia Hameed}  is  a Ph.D. student in the School  of Electrical,  Computer, Biomedical Engineering,  Southern Illinois University,  IL, USA.
\end{IEEEbiographynophoto}

\vspace{-14mm}

\begin{IEEEbiographynophoto}{Esa Ollila}
 is a Professor with the Dept. Information and Communications Engineering, Aalto University, Finland.
\end{IEEEbiographynophoto}

\vspace{-14mm}

\begin{IEEEbiographynophoto}
{\bf Sergiy A. Vorobyov} is a Professor with the Dept. Information and Communications Engineering, Aalto University, Finland. He is an IEEE Fellow. 
\end{IEEEbiographynophoto}

\end{document}

%% file: FR_3_Spectrumb_V2.pdf_tex
\begingroup%
  \makeatletter%
  \providecommand\color[2][]{%
    \errmessage{(Inkscape) Color is used for the text in Inkscape, but the package 'color.sty' is not loaded}%
    \renewcommand\color[2][]{}%
  }%
  \providecommand\transparent[1]{%
    \errmessage{(Inkscape) Transparency is used (non-zero) for the text in Inkscape, but the package 'transparent.sty' is not loaded}%
    \renewcommand\transparent[1]{}%
  }%
  \providecommand\rotatebox[2]{#2}%
  \newcommand*\fsize{\dimexpr\f@size pt\relax}%
  \newcommand*\lineheight[1]{\fontsize{\fsize}{#1\fsize}\selectfont}%
  \ifx\svgwidth\undefined%
    \setlength{\unitlength}{395.52571935bp}%
    \ifx\svgscale\undefined%
      \relax%
    \else%
      \setlength{\unitlength}{\unitlength * \real{\svgscale}}%
    \fi%
  \else%
    \setlength{\unitlength}{\svgwidth}%
  \fi%
  \global\let\svgwidth\undefined%
  \global\let\svgscale\undefined%
  \makeatother%
  \begin{picture}(1,0.46474575)%
    \lineheight{1}%
    \setlength\tabcolsep{0pt}%
    \put(0,0){\includegraphics[width=\unitlength,page=1]{FR_3_Spectrumb_V2.pdf}}%
    \put(0.59775978,0.27143813){\color[rgb]{0,0,0}\rotatebox{0.01212842}{\makebox(0,0)[lt]{\lineheight{1.25}\smash{\begin{tabular}[t]{l}\textbf{$\text{24.25 GHz}$ }\end{tabular}}}}}%
    \put(0.16902998,0.27204824){\color[rgb]{0,0,0}\rotatebox{0.01212842}{\makebox(0,0)[lt]{\lineheight{1.25}\smash{\begin{tabular}[t]{l}\textbf{$\text{7.125 GHz}$ }\end{tabular}}}}}%
    \put(0.81853718,0.00180678){\color[rgb]{0,0,0}\rotatebox{0.01212842}{\makebox(0,0)[lt]{\lineheight{1.25}\smash{\begin{tabular}[t]{l}\textbf{$\text{24.25 GHz}$ }\end{tabular}}}}}%
    \put(0,0){\includegraphics[width=\unitlength,page=2]{FR_3_Spectrumb_V2.pdf}}%
    \put(0.51223924,0.13984427){\color[rgb]{0,0,0}\makebox(0,0)[lt]{\lineheight{1.25}\smash{\begin{tabular}[t]{l}\textbf{$\textbf{Regions-1,2,3}$}\end{tabular}}}}%
    \put(0,0){\includegraphics[width=\unitlength,page=3]{FR_3_Spectrumb_V2.pdf}}%
    \put(0.20949669,0.13860811){\color[rgb]{0,0,0}\makebox(0,0)[lt]{\lineheight{1.25}\smash{\begin{tabular}[t]{l}\textbf{$\textbf{Region-1}$}\end{tabular}}}}%
    \put(0.22869584,0.03973398){\color[rgb]{0,0,0}\makebox(0,0)[lt]{\lineheight{1.25}\smash{\begin{tabular}[t]{l}\textbf{$\textbf{Regions--2,3}$}\end{tabular}}}}%
    \put(0.82516021,0.27076345){\color[rgb]{0,0,0}\rotatebox{0.02455267}{\makebox(0,0)[lt]{\lineheight{1.25}\smash{\begin{tabular}[t]{l}\textbf{$\text{71 GHz}$ }\end{tabular}}}}}%
    \put(0.03041931,0.00356189){\color[rgb]{0,0,0}\rotatebox{0.30033603}{\makebox(0,0)[lt]{\lineheight{1.25}\smash{\begin{tabular}[t]{l}\textbf{$\text{7.125 GHz}$ }\end{tabular}}}}}%
    \put(0.20728918,0.00240142){\color[rgb]{0,0,0}\rotatebox{0.19117612}{\makebox(0,0)[lt]{\lineheight{1.25}\smash{\begin{tabular}[t]{l}\textbf{$\text{7.75 GHz}$ }\end{tabular}}}}}%
    \put(0.315316,0.00235209){\color[rgb]{0,0,0}\rotatebox{0.23969987}{\makebox(0,0)[lt]{\lineheight{1.25}\smash{\begin{tabular}[t]{l}\textbf{$\text{8.4GHz}$ }\end{tabular}}}}}%
    \put(0.42277687,0.00176663){\color[rgb]{0,0,0}\rotatebox{0.24680869}{\makebox(0,0)[lt]{\lineheight{1.25}\smash{\begin{tabular}[t]{l}\textbf{$\text{14.8 GHz}$ }\end{tabular}}}}}%
    \put(0.58021189,0.00167402){\color[rgb]{0,0,0}\makebox(0,0)[lt]{\lineheight{1.25}\smash{\begin{tabular}[t]{l}\textbf{$\text{15.35 GHz}$ }\end{tabular}}}}%
    \put(0,0){\includegraphics[width=\unitlength,page=4]{FR_3_Spectrumb_V2.pdf}}%
    \put(0.11468793,0.21077961){\color[rgb]{0,0,0}\makebox(0,0)[lt]{\lineheight{1.25}\smash{\begin{tabular}[t]{l}\textbf{$\text{FR1}$}\end{tabular}}}}%
    \put(0.08545069,0.18670109){\color[rgb]{0,0,0}\makebox(0,0)[lt]{\lineheight{1.25}\smash{\begin{tabular}[t]{l}\textbf{$\text{Sub-6 GHz}$ }\end{tabular}}}}%
    \put(0,0){\includegraphics[width=\unitlength,page=5]{FR_3_Spectrumb_V2.pdf}}%
    \put(0.43648586,0.20903055){\color[rgb]{0,0,0}\makebox(0,0)[lt]{\lineheight{1.25}\smash{\begin{tabular}[t]{l}\textbf{$\text{FR3}$}\end{tabular}}}}%
    \put(0.39667109,0.18672398){\color[rgb]{0,0,0}\makebox(0,0)[lt]{\lineheight{1.25}\smash{\begin{tabular}[t]{l}\textbf{$\text{6G cmWaves}$}\end{tabular}}}}%
    \put(0.54955748,0.32761204){\color[rgb]{0,0,0}\makebox(0,0)[lt]{\lineheight{1.25}\smash{\begin{tabular}[t]{l}\textbf{$\text{K-band}$}\end{tabular}}}}%
    \put(0.63861619,0.32763674){\color[rgb]{0,0,0}\makebox(0,0)[lt]{\lineheight{1.25}\smash{\begin{tabular}[t]{l}\textbf{$\text{Ka-band}$}\end{tabular}}}}%
    \put(0.74003861,0.32689716){\color[rgb]{0,0,0}\makebox(0,0)[lt]{\lineheight{1.25}\smash{\begin{tabular}[t]{l}\textbf{$\text{U-band}$}\end{tabular}}}}%
    \put(0,0){\includegraphics[width=\unitlength,page=6]{FR_3_Spectrumb_V2.pdf}}%
    \put(0.72537224,0.20960155){\color[rgb]{0,0,0}\makebox(0,0)[lt]{\lineheight{1.25}\smash{\begin{tabular}[t]{l}\textbf{$\text{FR2}$}\end{tabular}}}}%
    \put(0.68841443,0.18762575){\color[rgb]{0,0,0}\makebox(0,0)[lt]{\lineheight{1.25}\smash{\begin{tabular}[t]{l}\textbf{$\text{5G mmWave}$}\end{tabular}}}}%
    \put(0.50353557,0.36891915){\color[rgb]{0,0,0}\makebox(0,0)[lt]{\lineheight{1.25}\smash{\begin{tabular}[t]{l}\textbf{$\text{18 GHz}$ }\end{tabular}}}}%
    \put(0.45075438,0.27222859){\color[rgb]{0,0,0}\makebox(0,0)[lt]{\lineheight{1.25}\smash{\begin{tabular}[t]{l}\textbf{$\text{18 GHz}$ }\end{tabular}}}}%
    \put(0.09923409,0.37058623){\color[rgb]{0,0,0}\makebox(0,0)[lt]{\lineheight{1.25}\smash{\begin{tabular}[t]{l}\textbf{$\text{3.3 GHz}$ }\end{tabular}}}}%
    \put(0.00224109,0.37285007){\color[rgb]{0,0,0}\makebox(0,0)[lt]{\lineheight{1.25}\smash{\begin{tabular}[t]{l}\textbf{$\text{410 MHz}$ }\end{tabular}}}}%
    \put(0.21959703,0.37013005){\color[rgb]{0,0,0}\makebox(0,0)[lt]{\lineheight{1.69000006}\smash{\begin{tabular}[t]{l}\textbf{$\text{3.8  GHz}$ }\end{tabular}}}}%
    \put(0.30607412,0.37066972){\color[rgb]{0,0,0}\makebox(0,0)[lt]{\lineheight{1.25}\smash{\begin{tabular}[t]{l}\textbf{$\text{6.4 GHz}$ }\end{tabular}}}}%
    \put(0.40897106,0.36895387){\color[rgb]{0,0,0}\makebox(0,0)[lt]{\lineheight{1.25}\smash{\begin{tabular}[t]{l}\textbf{$\text{7.1 GHz}$ }\end{tabular}}}}%
    \put(0.59285862,0.36897935){\color[rgb]{0,0,0}\makebox(0,0)[lt]{\lineheight{1.25}\smash{\begin{tabular}[t]{l}\textbf{$\text{26.5 GHz}$ }\end{tabular}}}}%
    \put(0.69371003,0.37003529){\color[rgb]{0,0,0}\makebox(0,0)[lt]{\lineheight{1.25}\smash{\begin{tabular}[t]{l}\textbf{$\text{40 GHz}$ }\end{tabular}}}}%
    \put(0.77885474,0.36897924){\color[rgb]{0,0,0}\makebox(0,0)[lt]{\lineheight{1.25}\smash{\begin{tabular}[t]{l}\textbf{$\text{50 GHz}$ }\end{tabular}}}}%
    \put(0.8625828,0.36950065){\color[rgb]{0,0,0}\makebox(0,0)[lt]{\lineheight{1.25}\smash{\begin{tabular}[t]{l}\textbf{$\text{75 GHz}$ }\end{tabular}}}}%
    \put(0,0){\includegraphics[width=\unitlength,page=7]{FR_3_Spectrumb_V2.pdf}}%
    \put(0.83499681,0.3271842){\color[rgb]{0,0,0}\makebox(0,0)[lt]{\lineheight{1.25}\smash{\begin{tabular}[t]{l}\textbf{$\text{V-band}$}\end{tabular}}}}%
    \put(0,0){\includegraphics[width=\unitlength,page=8]{FR_3_Spectrumb_V2.pdf}}%
    \put(0.32304047,0.32555391){\color[rgb]{0,0,0}\makebox(0,0)[lt]{\lineheight{1.25}\smash{\begin{tabular}[t]{l}\textbf{$\text{ New 5G-band}$}\end{tabular}}}}%
    \put(0,0){\includegraphics[width=\unitlength,page=9]{FR_3_Spectrumb_V2.pdf}}%
    \put(0.10946374,0.32683){\color[rgb]{0,0,0}\makebox(0,0)[lt]{\lineheight{1.25}\smash{\begin{tabular}[t]{l}\textbf{$\text{ Current 5G-band}$}\end{tabular}}}}%
    \put(0,0){\includegraphics[width=\unitlength,page=10]{FR_3_Spectrumb_V2.pdf}}%
  \end{picture}%
\endgroup%

%% file: system_model_FR3_Kavindu_2.pdf_tex
\begingroup%
  \makeatletter%
  \providecommand\color[2][]{%
    \errmessage{(Inkscape) Color is used for the text in Inkscape, but the package 'color.sty' is not loaded}%
    \renewcommand\color[2][]{}%
  }%
  \providecommand\transparent[1]{%
    \errmessage{(Inkscape) Transparency is used (non-zero) for the text in Inkscape, but the package 'transparent.sty' is not loaded}%
    \renewcommand\transparent[1]{}%
  }%
  \providecommand\rotatebox[2]{#2}%
  \newcommand*\fsize{\dimexpr\f@size pt\relax}%
  \newcommand*\lineheight[1]{\fontsize{\fsize}{#1\fsize}\selectfont}%
  \ifx\svgwidth\undefined%
    \setlength{\unitlength}{359.6033676bp}%
    \ifx\svgscale\undefined%
      \relax%
    \else%
      \setlength{\unitlength}{\unitlength * \real{\svgscale}}%
    \fi%
  \else%
    \setlength{\unitlength}{\svgwidth}%
  \fi%
  \global\let\svgwidth\undefined%
  \global\let\svgscale\undefined%
  \makeatother%
  \begin{picture}(1,0.62243138)%
    \lineheight{1}%
    \setlength\tabcolsep{0pt}%
    \put(0,0){\includegraphics[width=\unitlength,page=1]{system_model_FR3_Kavindu_2.pdf}}%
    \put(0.16790238,0.54394552){\color[rgb]{0,0,0}\makebox(0,0)[lt]{\lineheight{1.25}\smash{\begin{tabular}[t]{l}Satellite-1 \end{tabular}}}}%
    \put(0.45377416,0.5688436){\color[rgb]{0,0,0}\makebox(0,0)[lt]{\lineheight{1.25}\smash{\begin{tabular}[t]{l}Satellite-J \end{tabular}}}}%
    \put(0.14537748,0.23181524){\color[rgb]{0,0,0}\makebox(0,0)[lt]{\lineheight{1.25}\smash{\begin{tabular}[t]{l}UE-1\end{tabular}}}}%
    \put(0.03062465,0.28431422){\color[rgb]{0,0,0}\makebox(0,0)[lt]{\lineheight{1.25}\smash{\begin{tabular}[t]{l}Clutter-$1$\end{tabular}}}}%
    \put(0.37972037,0.0606282){\color[rgb]{0,0,0}\makebox(0,0)[lt]{\lineheight{1.25}\smash{\begin{tabular}[t]{l}Clutter-$c$\end{tabular}}}}%
    \put(0.67651056,0.26387954){\color[rgb]{0,0,0}\makebox(0,0)[lt]{\lineheight{1.25}\smash{\begin{tabular}[t]{l}Clutter-$N_c$\end{tabular}}}}%
    \put(0.32859604,0.12581818){\color[rgb]{0,0,0}\makebox(0,0)[lt]{\lineheight{1.25}\smash{\begin{tabular}[t]{l}UE-$k$\end{tabular}}}}%
    \put(0.70788175,0.19228843){\color[rgb]{0,0,0}\makebox(0,0)[lt]{\lineheight{1.25}\smash{\begin{tabular}[t]{l}UE-$K$\end{tabular}}}}%
    \put(0.0224466,0.10236377){\color[rgb]{0,0,0}\makebox(0,0)[lt]{\lineheight{1.25}\smash{\begin{tabular}[t]{l}Extended target \end{tabular}}}}%
    \put(0.00356009,0.06470095){\color[rgb]{0,0,0}\makebox(0,0)[lt]{\lineheight{1.25}\smash{\begin{tabular}[t]{l}with $N_s$ scatterers\end{tabular}}}}%
    \put(0.35459641,0.23074003){\color[rgb]{0,0,0}\makebox(0,0)[lt]{\lineheight{1.25}\smash{\begin{tabular}[t]{l}\textbf{ISAC-enabled}\end{tabular}}}}%
    \put(0.34784431,0.19647768){\color[rgb]{0,0,0}\makebox(0,0)[lt]{\lineheight{1.25}\smash{\begin{tabular}[t]{l}\textbf{BS with ELAA}\end{tabular}}}}%
    \put(0,0){\includegraphics[width=\unitlength,page=2]{system_model_FR3_Kavindu_2.pdf}}%
    \put(0.52761118,0.46217392){\color[rgb]{0,0,0}\makebox(0,0)[lt]{\lineheight{1.25}\smash{\begin{tabular}[t]{l}Rx array\end{tabular}}}}%
    \put(0.52719173,0.43693367){\color[rgb]{0,0,0}\makebox(0,0)[lt]{\lineheight{1.25}\smash{\begin{tabular}[t]{l}Tx array\end{tabular}}}}%
    \put(0,0){\includegraphics[width=\unitlength,page=3]{system_model_FR3_Kavindu_2.pdf}}%
    \put(0.17906494,0.4545869){\color[rgb]{0,0,0}\makebox(0,0)[lt]{\lineheight{1.25}\smash{\begin{tabular}[t]{l}CPE\end{tabular}}}}%
    \put(0.4660936,0.13894508){\color[rgb]{0,0,0}\makebox(0,0)[lt]{\lineheight{1.25}\smash{\begin{tabular}[t]{l}CPE\end{tabular}}}}%
    \put(0.81743003,0.37686462){\color[rgb]{0,0,0}\makebox(0,0)[lt]{\lineheight{1.25}\smash{\begin{tabular}[t]{l}CPE\end{tabular}}}}%
    \put(0.76203478,0.56507549){\color[rgb]{0,0,0}\makebox(0,0)[lt]{\lineheight{1.25}\smash{\begin{tabular}[t]{l}Omni-directional\end{tabular}}}}%
    \put(0.71874092,0.14952144){\color[rgb]{0,0,0}\makebox(0,0)[lt]{\lineheight{1.25}\smash{\begin{tabular}[t]{l}Commun. beams\end{tabular}}}}%
    \put(0.71584667,0.1190286){\color[rgb]{0,0,0}\makebox(0,0)[lt]{\lineheight{1.25}\smash{\begin{tabular}[t]{l}Sensing beams\end{tabular}}}}%
    \put(0.71595403,0.08766825){\color[rgb]{0,0,0}\makebox(0,0)[lt]{\lineheight{1.25}\smash{\begin{tabular}[t]{l}Sattelite-CPE beams\end{tabular}}}}%
    \put(0.71616403,0.05367128){\color[rgb]{0,0,0}\makebox(0,0)[lt]{\lineheight{1.25}\smash{\begin{tabular}[t]{l}CPE beacons\end{tabular}}}}%
    \put(0.80615099,0.53761185){\color[rgb]{0,0,0}\makebox(0,0)[lt]{\lineheight{1.25}\smash{\begin{tabular}[t]{l}beacon for\end{tabular}}}}%
    \put(0.76302059,0.5093034){\color[rgb]{0,0,0}\makebox(0,0)[lt]{\lineheight{1.25}\smash{\begin{tabular}[t]{l}CPE sensing and \end{tabular}}}}%
    \put(0.80262542,0.48023001){\color[rgb]{0,0,0}\makebox(0,0)[lt]{\lineheight{1.25}\smash{\begin{tabular}[t]{l}localization\end{tabular}}}}%
  \end{picture}%
\endgroup%

%% file: 1main_CameraReady.bbl
\begin{thebibliography}{10}
	
	\bibitem{Liu2022}
	F.~Liu \emph{et~al.}, ``{Integrated Sensing and Communications: Toward
		Dual-Functional Wireless Networks for {6G} and Beyond},'' \emph{{IEEE} J.
		Sel. Areas Commun.}, vol.~40, no.~6, pp. 1728--1767, 2022.
	
	\bibitem{Liu2018}
	------, ``{Toward Dual-Functional Radar-Communication Systems: Optimal Waveform
		Design},'' \emph{{IEEE} Trans. Signal Process.}, vol.~66, no.~16, pp.
	4264--4279, 2018.
	
	\bibitem{vaezi2025ai}
	M.~Vaezi \emph{et~al.}, ``{A Tutorial on {AI}-Empowered Integrated Sensing and
		Communications},'' \emph{{IEEE} Commun. Surv. Tutor.},    2026 (Early Access).
	
	\bibitem{FCC2023}
	{FCC}, ``{A Preliminary View of Spectrum Bands in the 7.125-24 {GHz} Range; and
		a Summary of Spectrum Sharing Frameworks},'' \emph{{Tech. Adv. Counc. Adv.
			Spectr. Sharing Work. Group}}, pp. 1--29, 2023.
	
	\bibitem{Bjornson2024}
	E.~Bj{\"o}rnson \emph{et~al.}, ``{Enabling {6G} Performance in the Upper
		Mid-Band by Transitioning From Massive to Gigantic {MIMO}},'' \emph{IEEE Open
		J. Commun. Soc}, vol.~6, pp. 5450--5463, 2025.
	
	\bibitem{Kang2024}
	S.~Kang \emph{et~al.}, ``{Cellular Wireless Networks in the Upper Mid-Band},''
	\emph{IEEE Open J. Commun. Soc.}, vol.~5, pp. 2058--2075, 2024.
	
	\bibitem{Bazzi2025}
	A.~Bazzi \emph{et~al.}, ``{Upper Mid-Band Spectrum for 6G: Vision, Opportunity
		and Challenges},'' \emph{IEEE Commun. Mag.}, vol.~64, no.~1, pp. 206--212,
	2026.
	
	\bibitem{Cui2023}
	Z.~Cui, P.~Zhang, and S.~Pollin, ``{6G Wireless Communications in 7–24 GHz
		Band: Opportunities, Techniques, and Challenges},'' in \emph{Proc. IEEE Int.
		Symp. Dyn. Spectr. Access Netw. (DySPAN)}, 2025, pp. 1--8.
	
	\bibitem{Miao2025}
	H.~Miao \emph{et~al.}, ``{{6G} New Mid-Band/{FR3} (6–24 {GHz}): {C}hannel
		Measurement, Characteristics and Modeling},'' \emph{IEEE Open J. Commun.
		Soc.}, vol.~6, pp. 9942--9960, 2025.
	
	\bibitem{Nokia2024}
	Nokia, ``{Coverage evaluation of 7–15 {GHz} bands from existing sites},''
	\emph{White Paper}, pp. 1--15, 2024.
	
	\bibitem{Ramezani2024}
	P.~Ramezani and E.~Bj{\"o}rnson, ``{Near-Field Beamforming and Multiplexing
		Using Extremely Large Aperture Arrays},'' \emph{{Fundamentals of 6G
			Communications and Networking}}, pp. 317--349, 2024.
	
	\bibitem{Gao2015}
	X.~Gao \emph{et~al.}, ``{Massive {MIMO} Performance Evaluation Based on
		Measured Propagation Data},'' \emph{{IEEE} Trans. Wireless Commun.}, vol.~14,
	no.~7, pp. 3899--3911, 2015.
	
	\bibitem{Knott1993}
	K.~Eugene \emph{et~al.}, \emph{Radar Cross Section}.\hskip 1em plus 0.5em minus
	0.4em\relax Artech House, 1993.
	
	\bibitem{Zhao2014}
	B.~Zhao \emph{et~al.}, ``{Modeling and Analysis of Near-Field ISAC},''
	\emph{IEEE J. Sel. Top. Signal Process.}, vol.~18, no.~4, pp. 678--693, 2024.
	
	\bibitem{Khawar2015}
	A.~Khawar \emph{et~al.}, ``{Target Detection Performance of Spectrum Sharing
		{MIMO} Radars},'' \emph{IEEE Sens. J.}, vol.~15, no.~9, pp. 4928--4940, 2015.
	
\end{thebibliography}
